# An analysis of Twitter messages in the 2011 Tohoku Earthquake

Son Doan, Bao-Khanh Ho Vo, and Nigel Collier

National Institute of Informatics
Chiyoda-ku, Hitotsubashi, Tokyo, Japan
`{doan,khanhvo,collier}@nii.ac.jp`

**Abstract.** Social media such as Facebook and Twitter have proven to be a useful resource to understand public opinion towards real world events. In this paper, we investigate over 1.5 million Twitter messages (*tweets*) for the period 9th March 2011 to 31st May 2011 in order to track awareness and anxiety levels in the Tokyo metropolitan district to the 2011 Tohoku Earthquake and subsequent tsunami and nuclear emergencies. These three events were tracked using both English and Japanese tweets. Preliminary results indicated: 1) close correspondence between Twitter data and earthquake events, 2) strong correlation between English and Japanese tweets on the same events, 3) tweets in the native language play an important roles in early warning, 4) tweets showed how quickly Japanese people's anxiety returned to normal levels after the earthquake event. Several distinctions between English and Japanese tweets on earthquake events are also discussed. The results suggest that Twitter data can be used as a useful resource for tracking the public mood of populations affected by natural disasters as well as an early warning system.

## 1 Introduction

Social media such as Facebook and Twitter have proven to be useful resources for understanding public opinion towards natural disaster events. Such resources can be used to detect general events in politics, e.g., elections [9], and finance, e.g., stock market changes [2,9] and oil price changes [9], as well as in alerting disasters such as earthquakes and typhoons [11]. Other social data such as search queries have been successfully used in public health to build bio-surveillance systems for early warning of influenza-like illness [7][10], showing high correlations with Centers for Disease Control and Prevention (CDC) reports. Within the wider Web, the BioCaster project has worked on detecting and tracking infectious diseases using newswire reports [5]. Twitter, the largest micro-blogging service with about 200 milion users as of March 2011 [1], can generate 200 million *tweets* a day. Tweets are short but condensed personal messages with a 140 character limit designed for rapid reporting from mobile devices. Several applications using Twitter messages in biosurveillance systems have been developed. For example, Flu Detector used Twitter messages to detect ILI rate in United Kingdom [8], DIZIE, which is part of the BioCaster project, is an experiemental syndromic surveillance system [6].

The great Tohoku earthquake happenned on 11th March 2011 was the most powerful known earthquake to have hit Japan, and one of the five most powerful earthquakes in the world overall since modern record-keeping began in 1900 [3]. The earthquake triggered a tsunami, causing massive loss of life and destruction of infrastructure, and in turn leading to a number of nuclear accidents in Fukushima prefecture, affecting hundreds of thousand of residents. This was described as "the toughest and the most difficult crisis for Japan" by the Japanese Prime Minister [4]. Greater understanding of social responses during such disaster periods should help metropolitan governments and public health agencies to gain greater insights for preparedness and response. Twitter data, being real time and large-scale, offers a unique insight into public opinions as the disaster develops.

In this paper, we analyzed over 1.5 million Twitter messages for the period starting $9^{th}$ March 2011 until $31^{st}$ May 2011 – the time when the main crisis happened - in order to review social attitudes during the time when the earthquake occurred. We focused on tracking keywords related to three main topics: earthquake/tsunami, radiation and public anxiety for the Twitter user population in the metropolitan Tokyo area; an area that experienced severe tremors, social anxiety and mild radiation but no major loss of life. To gain greater insights into differing attitudes between local and foreign residents we explicitly differentiated English and Japanese tweets. Our results show high correlations between Twitter data and real world events as well as how quickly Japanese people's anxiety returned to a stable level after the disasters. To the best of our knowledge, this is the first such study on Twitter data during the 2011 Tohoku earthquake.

## 2 Methods

### a Twitter corpus

We collected Twitter data for three months, starting from March 9th 2011 to May 31st 2011 using Twitter API (http://dev.twitter.com/) with the geolocation feature set to track messages originating within Tokyo. The resulting corpus had a total of 48,870 tweets in English and 1,611,753 tweets in Japanese. The details of tweets by dates in both English and Japanese are depicted in Figure 1.

### b Earthquake events and relevant keywords

Our empirical analysis focuses on the events during the 2011 Tohoku earthquake. Within the stream of Twitter messages we studied three indicators of public response: 1) earthquake and tsunami, 2) radiation caused by the Fukushima Daiichi plant's meltdown, and 3) public anxiety. The first two types of indicators are aimed at showing people's awareness of the earthquake, tsunami and radiation and the last indicators looks at how people in Tokyo are anxious about these events. Essentially, the events happened as the sequences as follows: The first is the earthquake occurring at 05:46:23 UTC on Friday, 11th March 2011. The second is the tsunami which happened after the earthquake a few minutes. The third is the nuclear explosion at the Fukushima Daiichi plant which the first explosion at reactor 1 happened at 6:36 UTC on 12th March 2011.

**Figure 1.** Tweet numbers by dates in English (left) and in Japanese (right).

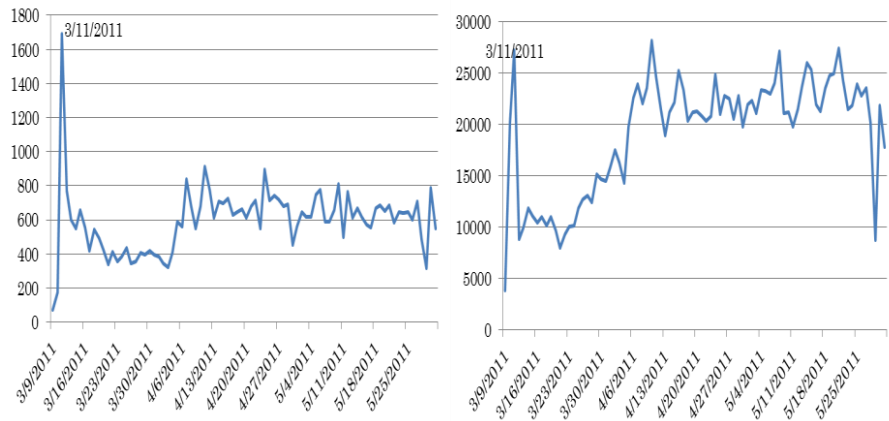

To being we manually investigated both English and Japanese tweets and constructed lists of key terms that are relevant to the events. The lists of English and Japanese terms indicating earthquake and tsunami, radiation and public anxiety are shown in Table 1. Although the number of terms might not be comprehensive we believed that they are good enough for our investigation. The reason we included earthquake and tsunami as a single event was because we found there were few English tweets about tsunami in our corpus.

c   **Data analysis**

Keyword filtering has shown to be a simple, but effective way to filter tweets for relevant topics [7,8,9,10]. In our study, we filtered tweets by event keywords in Table 1. We normalized tweets per day by dividing the total number of filtered tweets to the total number of tweets per day, i.e.,

$$f(\text{event}) = \frac{\#\text{filtered tweets per day}}{\#\text{tweets per day}}.$$

Where f(event) is the relative frequency of events per day.

**Table 1.** List of relevant keywords for the Earthquake and Tsunami, Radiation, and Anxiety events.

| English terms | Japanese terms |
|---|---|
| *Earthquake and Tsunami event* | |
| earthquake, quake, quaking, post-quake, shake, shaking, shock, aftershock, temblor, tremor, movement, sway, landslide seismic, seismography, seismometer, seismology, tsunami, wave | 大地震, 大震災<br>震災, 地震, 余震, 揺れ<br>震度, 震源, マグニチュード<br>津波 |
| *Radiation event* | |
| radiation, nuclear, reactor, radioactivity, radioactive, iodine, TEPCO, meltdown, explosion, power plant, micro sievert | 放射, 放射線, 放射能, 放射性物質, 原発, マイクロシーベルト, ヨウ素, イソジン, ヨウ化カリウム, 炉心溶融, メルトダウン |
| *Anxiety event* | |
| die, scary, scared, incredible, worrying, worried, anxious, annoying | 死亡, 死ぬ, やられてる, やばい, やばかった, ヤバい, やばっ, やべ, 怖い, 怖かった, 怖っ, すごい, すげえ, すげー, すっげー, びびる, びびった, 混乱, 微妙, 避難, 助けて, わかり辛い, 連絡とれない, 大変, 心配, 恐れ, 船酔いしそう |

## 3. Results and Discussions

Figure 1 supports Sakaki *et al.*'s [11] observations that the number of earthquake tweets increases significantly directly after a major earthquake. The data indicates that Twitter users would like to broadcast their experience immediately.

*Event 1*: Earthquake & tsunami event.
Earthquake and tsunami keyword frequencies for both English and Japanese are shown in Figure 2. We noticed that there is a sharp and sudden rise in the number of tweets immediately preceding the first major tremor. Note that the earthquake happened at 14:46:23 JST (05:46:23 UTC) on Friday, 11[th] March 2011 with 9.0 magnitude earthquake near the east coast of Honshu, Japan which was 373 km NE of Tokyo[1].

First, we considered how quickly Twitter users responded to the earthquake. It is unknown when the first public report about the earthquake was in Tokyo but the first tweet on the topic originating in Tokyo occurred at 05:48:08 UTC, 1 minute and 25 second right after the earthquake happened at the epicenter. It is unsurprising but noteworthy that the first tweet was in Japanese.

---

[1] http://earthquake.usgs.gov/earthquakes/recenteqsww/Quakes/usc0001xgp.php

Within our corpus the first English tweets on the earthquake are given below, with the first two tweets send from an iPhone:

```
2011-03-11T05:48:54   Huge earthquake in TK we are affected!
2011-03-11T05:49:01   BIG EARTHQUAKE!!!
2011-03-11T05:50:00   Massive quake in Tokyo
```

The first Japanese tweets on the earthquake are as follows.

```
2011-03-11T05:48:08   "地震！"              [Earthquake!]
2011-03-11T05:48:08   "地震だ〜縦揺れ！"      [Earthquake ~ vertical shake!]
2011-03-11T05:48:14   "地震！！！！"         [Earthquake!!!!]
```

We can easily see that first Twitter users responded very quickly, with the first English and Japanese tweets occurring about two minutes right after the earthquake happened. Japanese tweets preceded the English tweets by about 47 seconds. This might be because the numbers of Japanese language users are far greater than English language users in Tokyo. We note also that when the earthquake occurred, because of network outage there was no contact by cell phones but people could still access the Internet through 3G services with smartphones such as the iPhone.

We note that the first tweet from a Tokyo resident about a tsunami in Tohoku was a re-tweet at 06:02:35 UTC, 12 minutes after the first tsunami was reported. The first tweet about a tsunami was an eye witness tweet at 2011-03-11T 05:52:23 UTC, 6 minutes after the earthquake occurred at its epicentre.

```
2011-03-11T 05:52:23   "オレ、津波の様子見てくるわ！！！！！" [I can see tsunami is coming!!!!]
```

**Figure 2.** Keyword frequencies for the earthquake event over time for English (left) and Japanese tweets (right).

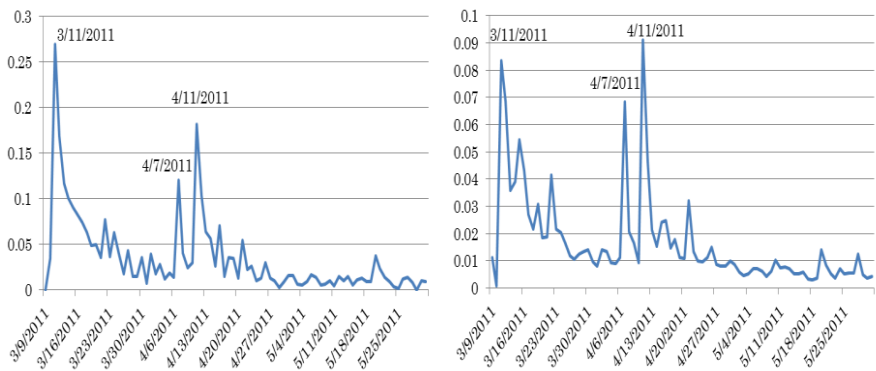

Let us consider the details of the aftershocks which were described in Wikipedia[2].

"Japan experienced over 900 aftershocks since the earthquake, with about 60 registering over magnitude 6.0 Mw and at least three over 7.0 Mw. A magnitude 7.7 Mw and a 7.9 Mw quake occurred on March 11 and *the third one struck offshore on 7 April with a disputed magnitude*"

---
[2] http://en.wikipedia.org/wiki/2011_Tōhoku_earthquake_and_tsunami

"Four days later *on April 11, another strong magnitude 6.6 Mw aftershock struck Fukushima,* causing additional damage and killing a total of three people"

As reported above, there are two other significant earthquake events: **7th April** and **11th April**. Both English and Japanese tweets in Figure 1 show how significant they are since peaks occur on both of those date with frequencies of 0.12 (English) and 0.07 (Japanese) at 7th April and 0.18 (English) and 0.09 (Japanese) at 11th April, respectively. It seems from our observations that Japanese language speakers were more concerned on 11th April than 7th April.

Below tweets were the first concerns about nuclear plants right after the earthquake.
    2011-03-11T05:57:53   "原発大丈夫かな？" [Is the nuclear power plant okay?]
    2011-03-11T08:43:36   "詳しく RT @u_tips: 福島第 2 原発が穏やかじゃないですね。" [In detail RT @u_tips: Fukushima Dai-ni power plant is on alert.]
    2011-03-11T09:50:49   "福島原発ヤバい状況らしい。。。政府が国民を欺かないことを願います" [The Fukushima plant is in a really bad situation… I hope that the government won't deceive the public.]

From drill down analysis we noticed that many people reported the situation happening in Tokyo from their own personal experiences such as a lack of food in convenience store on 11th and 12th March.
    2011-03-11T11:27:03     People r suggested to prepare an "emergency kit" consist of blanket, water, canned food, flashlight, aid kit, clothes #bigearthquakeinjapan
    2011-03-12T01:00:18      Wow. I've never seen a convenience store depleted of food before, even during the Great Handgun quake. At least I got toilet paper.

It is easy to see that such information could be automatically harvested for timely planning in future disasters. From Figure 2, we can see that both English and Japanese tweets correlate closely, reflecting the fact that public concern in both English and Japanese are the same during the earthquake events.

*Event 2*: Radiation event.
As reported in many public newswires, radiation was one of the main concerns after the earthquake and tsunami severely damaged the Fukushima Daiichi nuclear plant, causing three of its reactors to experience a meltdown although this fact was not confirmed until several months later. The radiation keyword frequencies for both English and Japanese tweets are depicted in Figure 3.

Following the timeline of the Fukushima Daiichi nuclear disaster[3], there were significant events about explosions of reactors as follows:

*12th March*, *15:36JST: Massive hydrogen explosion on the outer structure of the unit 1.*
*14th March*, *11:01JST: The unit 3 reactor building explodes, injuring six workers.*

---

[3] http://en.wikipedia.org/wiki/Timeline_of_the_Fukushima_I_nuclear_accidents

***15th March**, 20:00JST: A major part of the fuel in reactor 2 drops to the bottom of the reactor pressure vessel. Radiation levels at the plant rise significantly but subsequently fall back. Radiation equivalent dose rates of 400 millisieverts per hour (400 mSv/h) are observed at one location in the vicinity of unit 3.*

Figure 3 shows that after the March 11th earthquake, Japanese tweets showed further peaks on the 12th and 15th March whilst English tweets reached peaks one or two days later later on the 13th and 17th March, respectively. Although the cause is not clear this indicates that Japanese people in Tokyo were concerned about radiation earlier than foreign residents in Tokyo. Once again the results indicate the important role that aggregated tweets in the native languages play in early warnings.

When the earthquake hit the Fukushima nuclear plant on 11th April, both English and Japanese tweets reached their peak a day later, on 12th April. This indicates that the event is of major concern to both Japanese people and foreigner residents in Tokyo.

**Figure 3.** Keyword frequencies for the radiation event over time for English (left) and Japanese tweets (right).

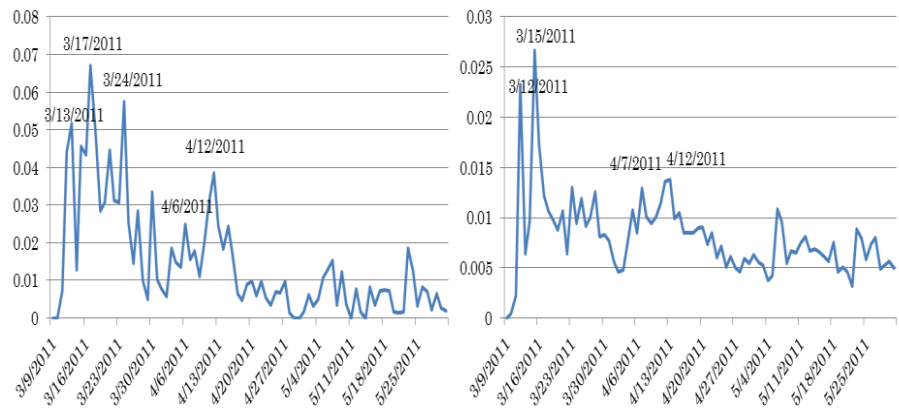

Below are some examples of early English tweets at the peak on 11th and 12th April.
```
  2011-04-11T23:21:09     Earthquakes, tsunamis, radiation what else you
got? Exploding Fuji? Let me get out my folding umbrella...  #fb
#quaketrashtalk #backchannel
  2011-04-11T23:51:09     This is the "New Normal" life after 3/11.
Always fearing how the situation of nuclear plant is #prayforjapan
  2011-04-12T00:08:01     Nuclear Agency Japan has increased the level
of nuclear disaster to level 7 as the worst, which equal to Chernobyl
#Fukushimasradiation
  2011-04-12T02:41:40     @<username> 4 of 6 reactors in meltdown. Spent
fuel rods melting. Two explosions, containment breached. Already a 7.
#fukushima #niisa
```

Topically, several tweets focus on concerns about radiation in relation to tap water in Tokyo since 13th March.
```
  2011-03-13T07:06:37     People are asked to close window, door; not to
use AC; use mask & not to drink tap water #Fukushimasradiation
```

```
    2011-03-23T06:18:50    210 becquerel iodine (normal 100 becquerel)
discovered in Tokyo tap water. Infants are urged to avoid drinking it
#Fukushimasradiation
```

*Event 3*: Anxiety event

Using keywords with Twitter data can track the public mood such as in elections or oil price changes [9]. It is interesting to see if they can be used to follow the anxiety of Tokyo residents during the earthquake. Figure 4 showed anxiety frequencies for English and Japanese tweets. Since the frequency of English tweets are relatively low since we had few tweets about anxiety, we included it here as a reference.

Figure 4 shows that the highest frequencies of the anxiety event were at the starting point -11$^{th}$ March and then decreased over a two week period and kept stable after that. There was a slight rise in concern on 11$^{th}$ April but the upward change was not so high when comparing to the earthquake and nuclear events in Japanese. This perhaps underlines the well reported response about how Japanese people kept calm during the disasters. Despite the relatively small number of English language tweets, aggregated message counts from foreign residents show many peaks. However, when comparing the peaks we can see similar trends with Japanese tweets, especially from 20$^{th}$ April to 11$^{th}$ May.

**Figure 4.** Frequencies of the anxiety event by dates of Japanese tweets.

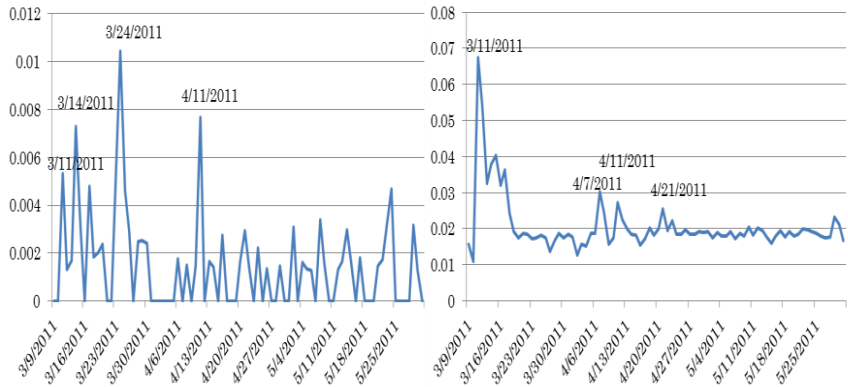

Below are some examples for anxiety of people in Tokyo in English tweets,
```
    2011-03-11T06:58:49    nerves frayed out on the streets. everyone
emptied out of buildings, on their phones, worried. some inspecting
damage to buildings
    2011-03-11T07:09:56    @<username> I'm okay, thx.  Worried about
others...
    2011-03-11T07:14:52    @<username> thanks ! Now Me and my family are
ok. But worrying about my men...
```

## 4    Conclusions

In this paper, we provided an empirical investigation that showed high correlations between aggregated tweets and disasters during earthquake events, i.e., earthquake and tsunami, radiation. It also appears to be potential for tracking public anxiety in resident populations affected by the disaster. The results reveal that tweets in the native language play an important role in early warning in terms of their volume and timeliness. Strong correlation between Twitter and public health events leads us to believe that Twitter data can be a useful resource in an early warning surveillance systems as well as a tool for analyzing public anxiety and needs during times of disaster.

In the future, we plan to extend our work on analysis to other aspects during the earthquake using publicly available metrics for evaluation. Automated methods to find relevant terms for tracking during disasters will also be investigated.

**Acknowledgements**
Authors would like to thank to Twitter Inc. for providing API functions to access the Twitter data.